# Quality Atomic Resolution Scanning Tunneling Microscope Imaging up to 27 T in Water-cooled Magnet

Wenjie Meng, [1, 2] Ying Guo, [1, 2] Yubin Hou, [1] and Qingyou Lu [1, 2, 3, a)]

[1] *High Magnetic Field Laboratory, Chinese Academy of Sciences and University of Science and Technology of China, Hefei, Anhui 230026, People's Republic of China*

[2] *Hefei National Laboratory for Physical Sciences at the Microscale, University of Science and Technology of China, Hefei, Anhui 230026, People's Republic of China*

[3] *Collaborative Innovation Center for Artificial Microstructure and Quantum Control, Nanjing 210093, People's Republic of China.*

[a)] *Author to whom correspondence should be addressed. Electronic mail: qxl@ustc.edu.cn. Tel.: 86-551-6360-0247.*

We report the achievement of the first atomically resolved scanning tunneling microscope (STM) imaging in a water-cooled magnet (WM), where the extremely harsh vibrations and noises have been the major challenge. This homebuilt WM-STM features an ultra-rigid and compact scan head in which the coarse approach is driven by our new design of the TunaDrive piezoelectric motor. A three-level spring hanging system is exploited for vibration isolation. Room-temperature raw-data images of graphite with quality atomic resolution were obtained in very high magnetic fields up to 27 T in a 32 mm bore WM whose absolute maximum field is 27.5 T at the power rating of 10 MW. This record of 27 T has exceeded the maximum field strength of the conventional superconducting magnets. Besides, our WM-STM has also paved the way to the STM imaging in the 45 T, 32 mm bore hybrid magnet, which is the world's flagship magnet and can produces the highest steady magnetic field at present.

## 1. Introduction

The scanning tunneling microscope (STM) is capable of providing information on both surface structures and local electronic density of states with atomic resolution, which has resulted in wide fundamental research applications.[1] It is nevertheless very sensitive to even weak sound and vibration disturbances and a carefully designed sound and vibration isolation system is typically needed.[2] One of the most important applications of the STM is to image in high magnetic field, which is crucial in the studies of high-temperature superconductors,[3-6] the dirac nature of the surface states of topological insulators,[7-9] the quantum hall effect in low dimension materials[10, 11] and the vortex formation in quantum dots,[12] etc. To this end, the STM is routinely housed in a superconducting magnet, which has the superiority of tranquility.[13-16] However, obtaining a magnetic field higher than 23.5 T from a superconducting magnet is very difficult due to the limitation of the critical current.[17] Therefore, the current highest magnetic field record for the STM is still no more than 18 T after so many years of developments.[13, 18] A water-cooled magnet (WM) or a hybrid magnet (which consists of a WM inside a superconducting magnet) can produce higher magnetic field, but the ultra-strong flow of the cooling water can also generate enormous vibration and sound, which is a huge challenge for the STM. Up to now, no atomically resolved STM imaging in a water-cooled magnet has been reported.

We have long been working on the development of harsh-condition STM and invented a series of high rigidity and compactness piezoelectric motors including the GeckoDrive,[19] TunaDriver,[20, 21] PandaDrive[19] and SpiderDrive,[22] which are to some extent immune to external vibrations.[19-24] Based on these systematic studies, the major challenge of achieving atomic resolution in the WM's ultra-harsh conditions has been overcome and breakthrough has consequently been made. Room-temperature raw-data images of graphite with excellent quality atomic resolution were obtained in very high magnetic fields up to 27 T in a 32 mm bore water-cooled magnet, which has exceeded the maximum field of the conventional superconducting magnets. Actually, our homebuilt WM-STM was so stable that good atomic resolution images could be obtained continuously without the need of any adjustment when the field changed from 27 T to zero.

Apparently, this WM-STM has cleared the way for STM imaging in the 45 T, 32 mm bore hybrid magnet, which is available at the National High Magnetic Field Laboratory (MagLab) of the U.S. and under construction (close to finish though) at the High Magnetic Field Lab, Chinese Academy of Sciences (CHMFL). This is the world's flagship magnet, which can achieve the strongest continuous manmade magnetic field on earth at present. It is now foreseeable that the highest magnetic field for atomic-resolution imaging will only be limited by the magnet itself. The WM-STM features an ultra rigid and compact scan head in which the coarse approach is driven by our new design of the TunaDrive piezoelectric motor.[20, 21] A three-level spring hanging system is exploited for vibration isolation. In this paper, the detailed structure and performance of the WM-STM will be presented and discussed.

## 2. Water-cooled magnet

The WM we used was built in house by CHMFL and labeled WM4, which has a 32 mm diameter room temperature bore. The absolute maximum magnetic field it can safely produce is 27.5 T when working at its power rating of 10 MW. For the sake of magnet protection, we rarely allow the field to go beyond 27 T. This is why the maximum field was set to 27 T for the STM measurements in this paper. The length of the bore is 1299.7 mm and the location of the maximum field is 634.7 mm below the upper end of the bore. The WM cannot run continuously for more than 6 hours because the cooling deionized water will be used up in the storage tank.

## 3. STM head unit

Our WM-STM (see Fig. 1) mainly consists of three portions: the STM head unit, the vacuum chamber system and the vibration isolation system.

The main consideration in designing the STM head unit is compactness and rigidity, which can enhance its eigen-frequency[25] and reduce the entered disturbing energy.[26] To this end, we adopt a tiny EBL#3 type (from EBL Products Inc.) piezoelectric tube scanner (PTS) of 3.2 mm outer diameter, 8 mm length and 0.5 mm wall thickness, which is firmly glued by epoxy inside a polished sapphire tube of 5 mm outer diameter and 14 mm length. The sapphire tube is spring clamped on the two parallel knife edges of a tungsten rail piece, which is in turn encased in a cylindrical titanium base (see Fig. 2d).[27] To implement the tip-sample coarse approach, the sapphire tube needs to be pushed by a compact, yet very powerful piezoelectric motor.

We finally choose the TunaDrive[20] as the coarse approach motor. It is however redesigned to be made simpler, more compact and rigid as follows (see Fig. 2a and 2b): Two polished sapphire rods of 3.5 mm diameter and 50 mm length are mounted in parallel on the aforementioned cylindrical titanium base and in parallel with the PTS. These sapphire rods serve as the guiding rails between which three stacks of the driving piezo strips are spring clamped. Unlike the five-stack TunaDrive described in Ref. 21, we use only three stacks here: The central stack (CS) is sandwiched by two side stacks, SS1 and SS2. Each of the SS1 and SS2 is a stack of 3 piezoelectric strips (EBL#3, from EBL Products Inc.) of 28 mm length, 9 mm width and 0.5 mm thickness, whereas the CS is a stack of 4 such piezo strips. At one end, the CS and SS1 are glued, whereas at the other end, the CS and SS2 are glued. This Z-type folded three-stack driving part is then inserted and tightly spring clamed between the sapphire rod rails. The work principle of this three-stack TunaDrive is similar to that of the five-stack TunaDrive (see Fig. 2c)[21]: Fast expanding and contracting the SS1 periodically and slow expanding the CS (the SS2 in still) will reduce the friction force on the SS1 and cause the SS1 to move to follow the expansion of the CS; then fast expanding and contracting the SS2 periodically and slow contracting the CS (the SS1 in still) will reduce the friction force on the SS2 and cause the SS2 to move to follow the contraction of the CS. In the end, a step move is produced, which can be repeated to form stepping. Compared with the five-stack TunaDrive, this three-stack version is better in simplicity, compactness and rigidity and the full length of the CS contributes to the output force. Its measured output force is greater than 1.5 N. There is a pair of hooks between the pushing end of the piezo motor and the sapphire tube that carries the PTS. The hooks allow the piezo motor to push the PTS for coarse approach and withdraw it in the opposite direction as needed.

**4. Vacuum chamber system**

The vacuum chamber system consists of the main chamber and the tubular chamber (see Fig. 3). The cylindrical main chamber is made of 304 stainless steel, which is vertically hung above the WM via one spring. It measures 200 mm in outer diameter and 536 mm in height. Inside the main chamber is a coaxially top-mounted cold hollow cylinder (CHC) made of 304 stainless steel whose wall is also hollow and serves as the container to store a maximum of 6 L liquid nitrogen. The CHC functions as a cryogenic pump to maintain the vacuum of the main chamber around $10^{-5}$ Torr which is necessary to isolate the sound noise generated by the WM. The spaces inside and outside the CHC are connected and vacuum.

The CF flanges on the side wall of the main chamber are designed for the connections with the electrical feedthrough (where the preamplifier circuit box is attached), viewport, manipulators, vacuum gauge and the sample/tip preparation chamber (which allows the tip and sample to enter the main chamber and be transferred to the STM head unit).

The tubular chamber is made of titanium, which is highly nonmagnetic. It measures 1000 mm in length, 28 mm in outer diameter and 0.9 mm in wall thickness. It is connected to the center CF flange on the bottom of the main chamber in a coaxial manner and is inserted into the WM from the top. It is very important to make sure that the tubular chamber is not pushed to touch the wall of the WM bore by the magnetic field during any STM measurement. Otherwise, no atomic resolution images can be obtained for sure. This is why we use titanium to make the tubular chamber. We also hang a 25 kg weight underneath the tubular chamber so as to ensure that the tubular chamber is not tilted inside the WM bore.

To acquire vacuum, we used a TV81M turbo pump (from Varian Inc.) to pump the main chamber until $10^{-6}$ Torr was achieved. Liquid nitrogen was then filled into the hollow wall of the CHC, which could enhance the vacuum to $4.5 \times 10^{-8}$ Torr. After turning off the turbo pump, the vacuum could remain no worse than $10^{-5}$ Torr for 4 hours even if all the liquid nitrogen had evaporated for a certain period of time. This vacuum was necessary to prevent the loud sound of the cooling water from impacting the STM measurement. We actually measured the strength data of the acoustic noises near the WM and in a normal lab as well using a commercial noise meter (DT-8852 from CEM Company), which were 85 dB and 52 dB, respectively. Obviously, the former is much larger, which needs to be blocked by sufficiently high vacuum.

**5. Vibration isolation system**

There are internal and external vibration isolations, which are both important in achieving atomic resolution. The external vibration isolation between the STM chamber system and the WM is accomplished by two levels of hanging via springs (see Fig. 1). The main chamber is hung above the WM by a spring from an aluminum frame, which stands on an aluminum board. The board is in turn hung by 4 springs from two aluminum beams that are supported by two stacks of concrete bricks, respectively. Each concrete brick measures 850 mm in length, 450 mm in width, 150 mm in height and 100 kg in weight. There are rubber plates between the neighboring concrete bricks for further damping and vibration isolation. The two stacks of the concrete bricks are piled up vertically on the platform that surrounds the WM. Here, the board is the first vibration isolation stage and the main chamber is the second vibration isolation stage. The measured characteristic frequency of the external vibration isolation system was less than 3 Hz.

We also measured the vibration spectra of the WM, the board and the main chamber using a commercial vibration meter (BVM-100-2S-J from Beijing Vibration Measurement Company). The results are compared in Fig. 4. Obviously, the vibration strength of the main chamber was reduced by a factor of more than 70 compared with that of the WM.

Inside the main chamber, the STM head unit is hung underneath a tungsten rod of 20 mm diameter and 100 mm length for internal vibration isolation. The tungsten rod is in turn hung through a rotary wheel at the upper center of the main chamber, which is controlled by a rotary manipulator (see Fig. 3). By rotating the wheel forward or backward, the STM head unit can be either lowered into the tubular chamber to reach the field center of the WM for STM measurement, or lifted up into the main chamber for sample and tip replacements.

**6. Performance**

Figure 5 shows the constant height raw-data images of a highly oriented pyrolytic graphite (HOPG) sample taken in 27 T in WM4. They were scanned in room temperature with the scan sizes being $2.1 \times 2.1$ and $3.6 \times 3.6$ nm$^2$, respectively. Quality atomic resolution was repeatable at different scan sizes. The tunneling current ranges were from 247 to 333 nA and from 247 to 307 nA, respectively. The scan rate was 8 lines per second. The sample was biased at +100 mV. The tip was a cut Pt-Ir wire with the Pt to Ir ratio being 9:1.

To show the excellent stability and repeatability of the WM-STM better, we continuously took images of the same sample area while we were reducing the magnetic field from 27 T to 0 at a constant rate of 0.05 T per second. The image series for this procedure is shown in Fig. 6. Clear atomic resolution remained all the time with a total lateral drift being as low as 0.3 nm.

## 7. Conclusion

In brief, we have realized STM imaging in a water-cooled magnet for the first time and obtained clear atomic-resolution images in magnetic fields up to 27 T, which is much higher than the maximum field of the conventional superconducting magnets. This achievement provides new technology for the strong magnetic field STM. It also makes ready for us to implement the STM imaging in the 45 T hybrid magnet of CHMFL whose bore size is the same as the WM we used in this work. This hybrid magnet will be up for use soon.

## 8. Acknowledgments

This work was supported by the Major / Innovative Program of Development Foundation of Hefei Center for Physical Science and Technology, the Project of the Chinese National High Magnetic Field Facilities, and the National Natural Science Foundation of China under Grant No. U1232210, 11204306 and 11374278.

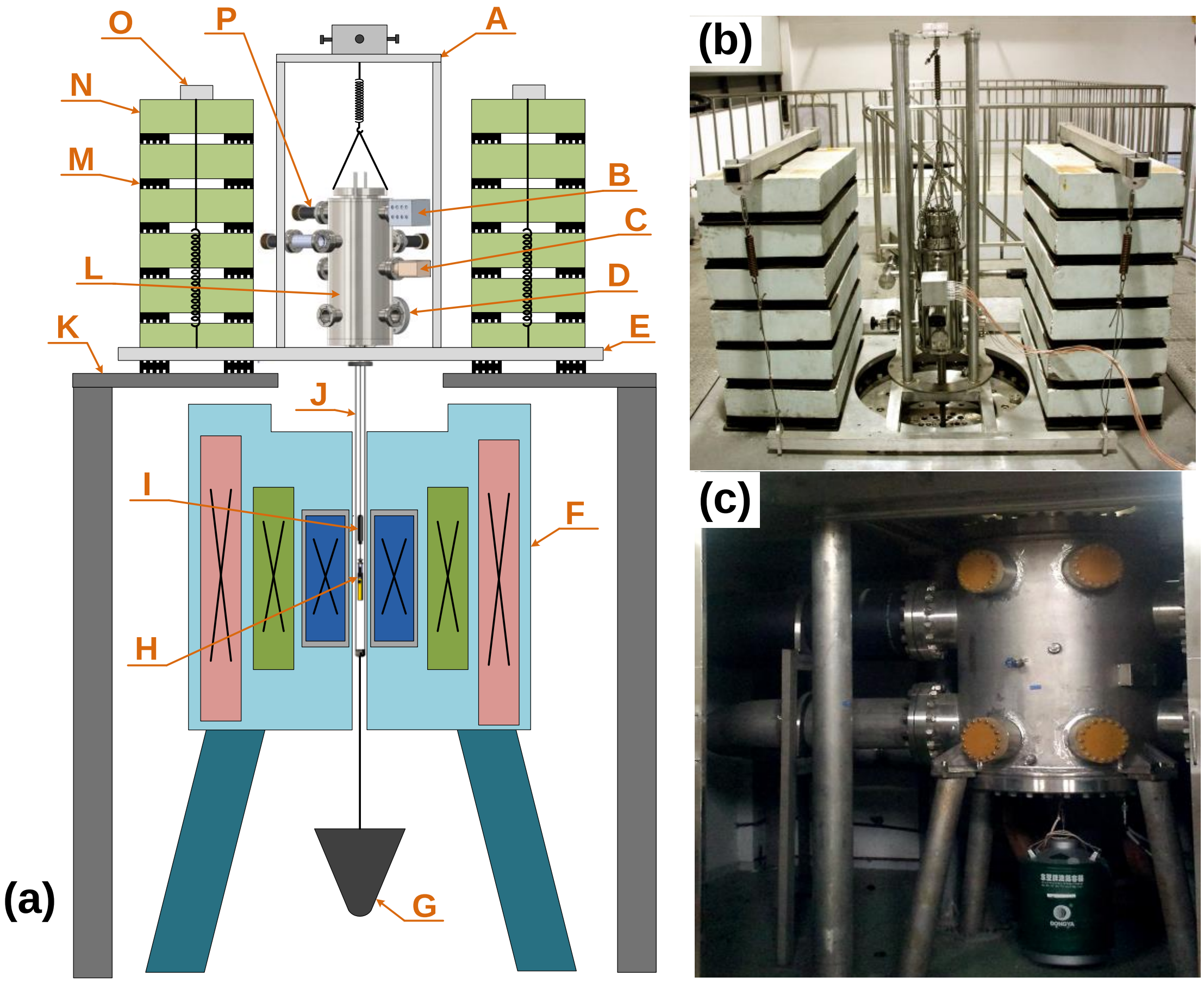

FIG. 1. (a) Schematic drawing of the WM-STM system: A. aluminum frame, B. preamplifier circuit box (which is attached to the electronic feedthrough), C. vacuum gauge, D. viewport, E. aluminum board, F. water-cooled magnet, G. weight, H. STM head unit, I. tungsten rod, J. tubular chamber, K. platform, L. main chamber, M. rubber plate, N. concrete brick, O. aluminum beam, P. rotary manipulator. (b) Photograph of the WM-STM system with the external vibration isolation. (c) Photograph of the WM used with the 25 kg weight (which is actually a liquid nitrogen dewar) underneath the tubular chamber so as to ensure that the tubular chamber is not tilted inside the WM bore.

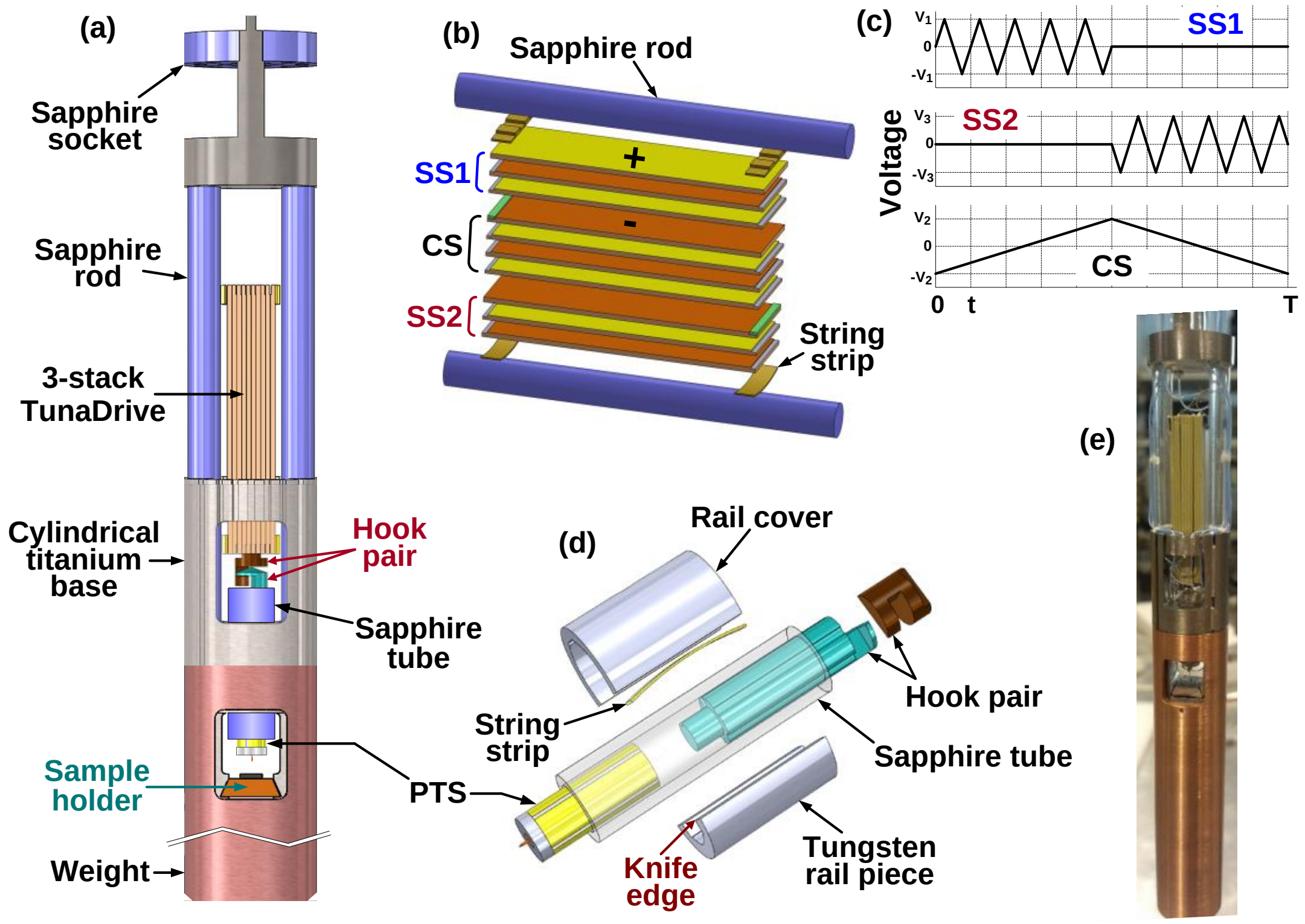

FIG. 2. (a) Schematic of the STM head unit. (b) Schematic drawing of the three-stack TunaDrive coarse approach motor. (c) Driving signals of the TunaDrive. (d) Exploded view of how the PTS is spring clamped onto the tungsten rail piece. (e) Photograph of the STM head unit.

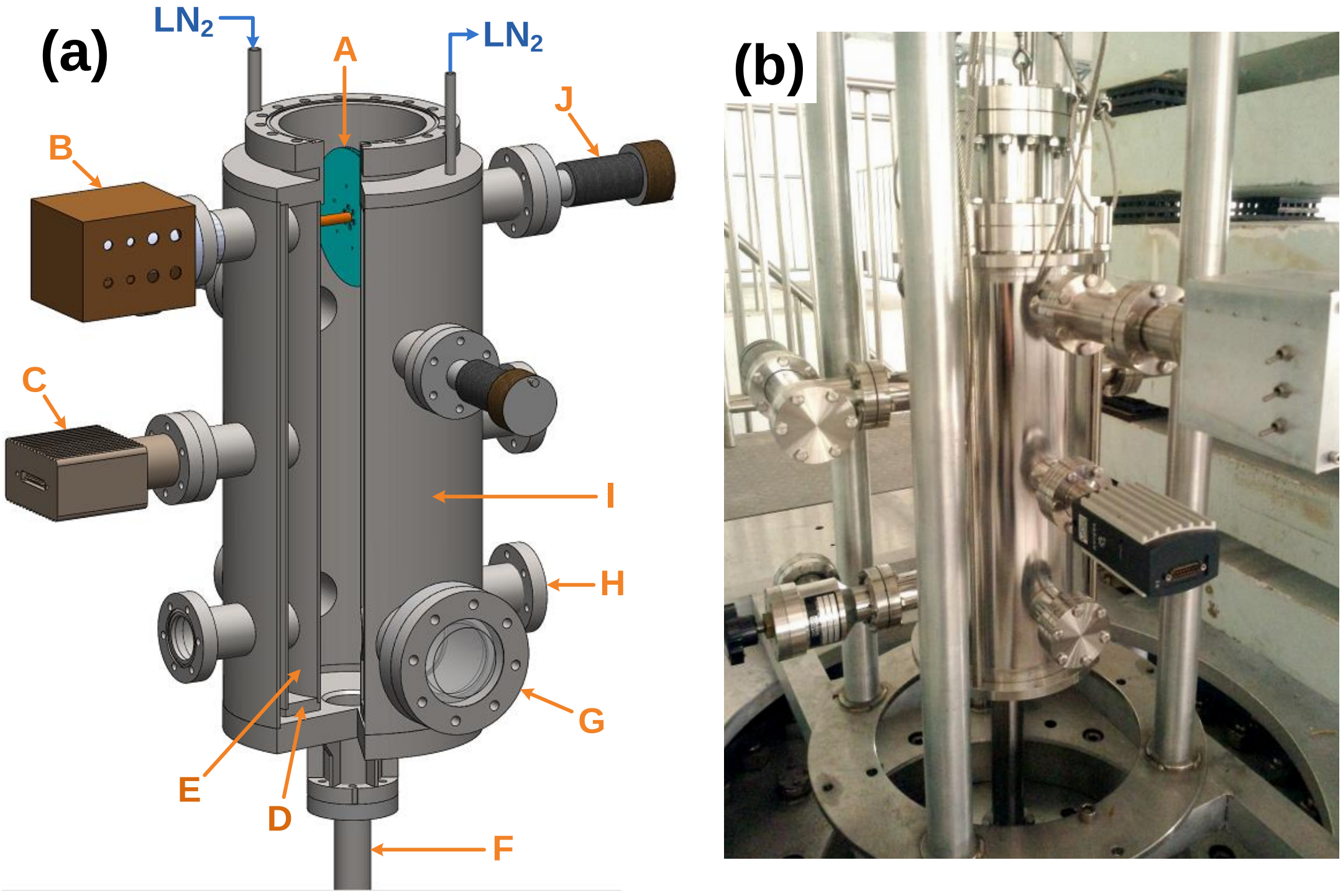

FIG. 3. (a) Structure diagram of the main chamber: A. rotary wheel, B. preamplifier circuit box (which is attached to the electronic feedthrough), C. vacuum gauge, D. cold hollow cylinder (CHC), E. hollow wall of the CHC, F. tubular chamber (also see fig. 1), G. viewport, H. flange for the sample/tip preparation chamber, I. main chamber, J. rotary manipulator. (b) Photograph of the main chamber.

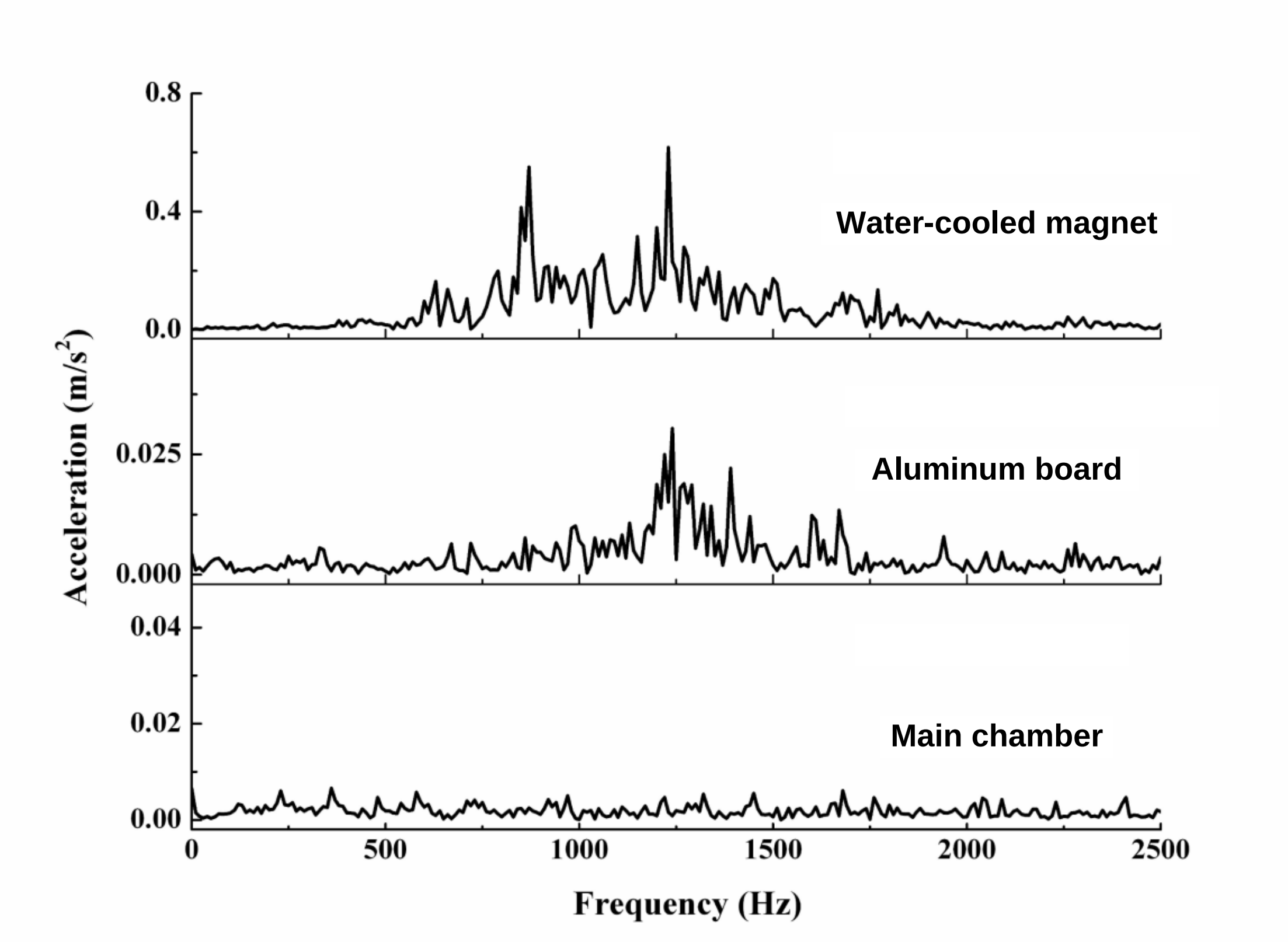

FIG. 4. Measured vibration spectra of the WM, the board and the main chamber.

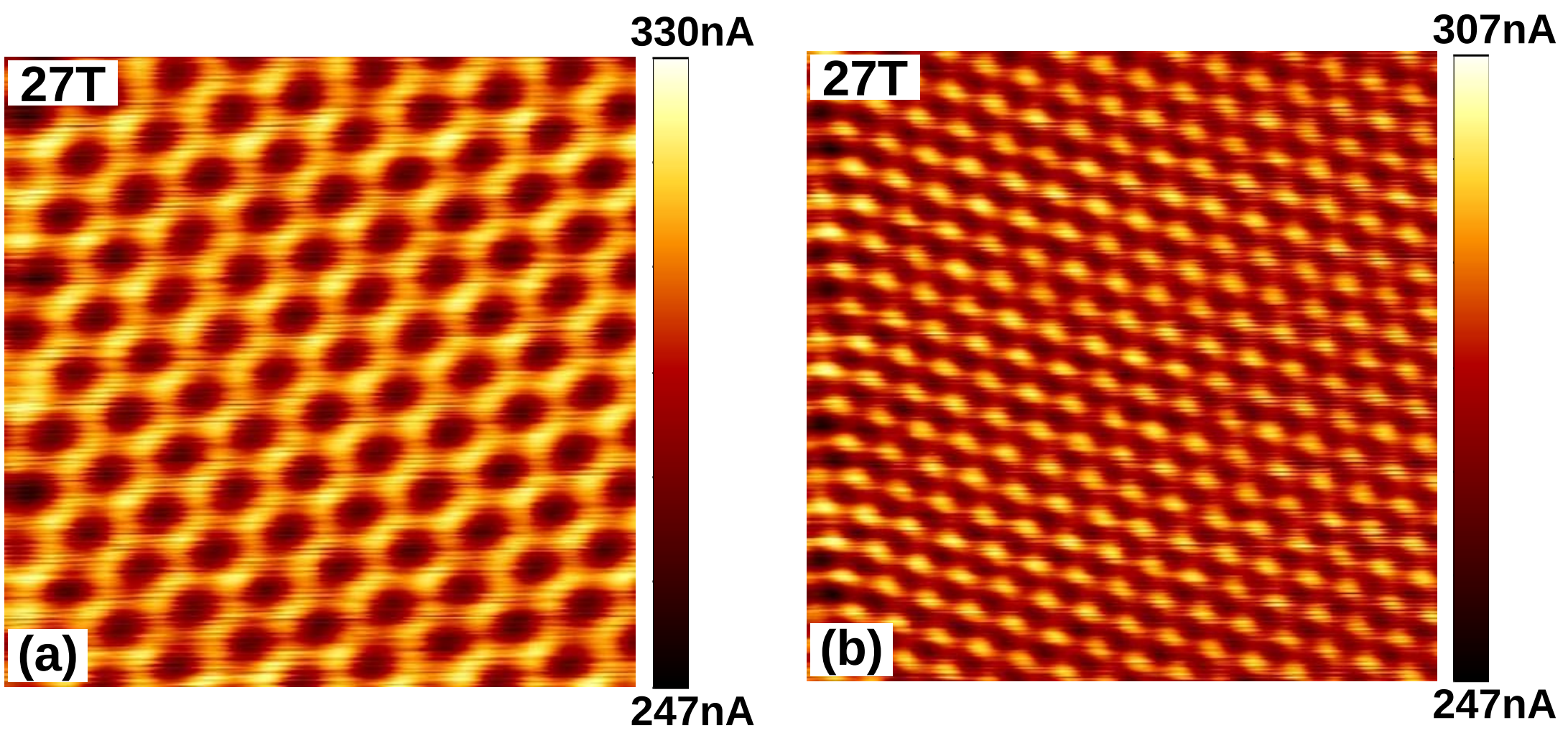

FIG. 5. Atomically resolved HOPG STM images (raw data) obtained in 27 T in WM4 under constant height mode and 100 mV sample positive bias. Scan sizes: (a) 2.1 $\times$ 2.1 $nm^2$ with the tunneling current range from 247 to 330 nA; (b) 3.6 $\times$ 3.6$nm^2$ with the tunneling current range from 247 to 307 nA.

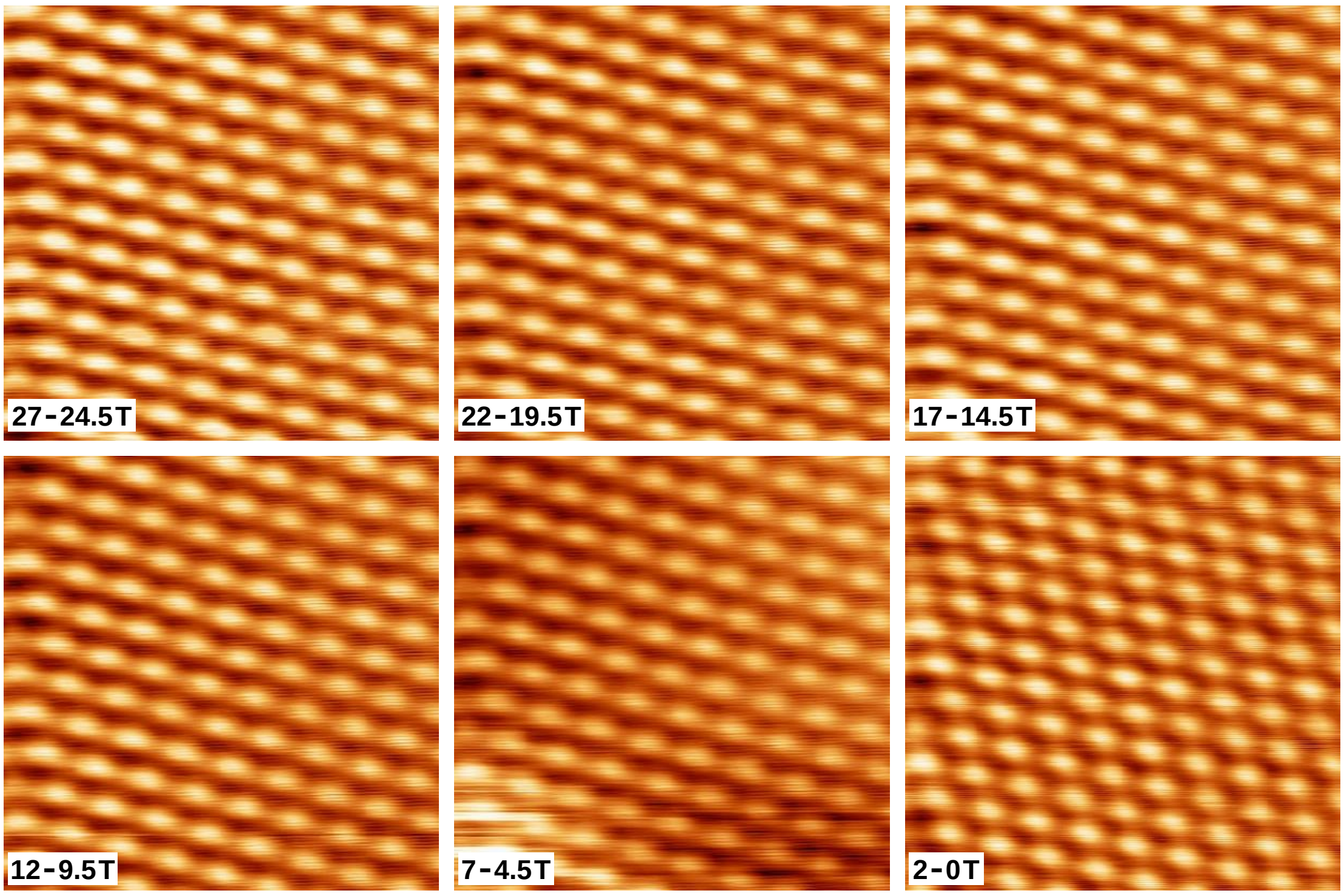

FIG. 6. Raw data image series of the same sample area taken continuously while the magnetic field was reducing from 27 T to 0 at a constant rate of 0.05 T/s. Image sizes are all 2.1 × 2.1 $nm^2$.